\documentclass{emulateapj}
\usepackage{apjfonts}

\newcommand{\hst}{{\it HST}}
\def\mso{M_\odot}

\def\lso{L_\odot}

\def\kms{\, {\rm km}\, {\rm s}^{-1}}

\slugcomment{To appear in ApJ Letters}

\shorttitle{Wind structure and rotation in NCC6543}
\shortauthors{Prinja et al.}

\begin{document}


\title{Looking deep into the Cat's Eye:  \\
    Structure and rotation in the fast wind of the PN central star of
NGC~6543\footnote{Based on observations made with the NASA/ESA Hubble Space
Telescope, obtained at the Space Telescope Science Institute, which is
operated by the Association of Universities for Research in Astronomy,
Inc., under NASA contract NAS 5-26555. These observations are associated
with program \# 12489.}}


\author{R.K. Prinja$^{1,}$, D.L. Massa$^{2}$, M. Cantiello$^{3}$}
\affil{$^1$ Dept. of Physics {\&} Astronomy, University College London,
Gower Street, London WC1E 6BT, UK}
\affil{$^2$ Space Telescope Science Institute, Baltimore, MD 21218, USA}
\affil{$^3$ Kavli Institute for Theoretical Physics, University of California Santa 
Barbara, CA 93106, USA}



\begin{abstract}
We present \hst/STIS time-series spectroscopy of the central star of the 
'Cat's Eye' planetary nebula NGC~6543.  Intensive monitoring of the UV 
lines over a 5.8 hour period reveals well defined details of large-scale 
structure in the fast wind, which are exploited to provide new constraints 
on the rotation rate of the central star.  We derive characteristics of the 
line profile variability that support a physical origin due to co-rotating 
interaction regions (CIRs) that are rooted at the stellar surface.  The 
recurrence time of the observed spectral signatures of the CIRs is used 
to estimate the rotation period of the central star and, adopting a radius 
between 0.3 and 0.6 $R_\odot$ constrains the rotational velocity to the 
range $54 \leq v_{rot} \leq 108$~$\kms$.  The implications of these 
results for single star evolution are discussed based on models calculated
here for low-mass stars.
Our models predict a
sub-surface convective layer in NGC~6543 which we argue to be
causally connected to the occurrence of structure in the fast wind.
\end{abstract}



\section{Introduction}
Planetary nebulae (PNe) display remarkably diverse morphologies, the large
majority of which are not spherical. The formation and shaping of bipolar
PNe, which can include axi- and point-symmetries, is currently a moot issue
in stellar astrophysics, with the likelihood that different mechanisms may
be more dominant in different nebulae.  An understanding of this PNe 
diversity encompasses the potential roles of several interesting physical 
processes, including magnetic fields, common envelope evolution in binary 
systems, stellar radiation fields and outflows.

One of contending scenarios for sculpting the nebulae is based on stellar
rotation in the Generalized Interacting Stellar Wind model \citep[GISW e.g.,][]{Frank1999}.
 In this case, asymmetry results from an equatorially enhanced
outflow from a post-asymptotic giant branch (AGB). The higher equatorial 
density constrains the nebular expansion in that region, leading to an
elliptical or bipolar PN. The origin of the requisite equatorial expansion
remains uncertain, but rotationally induced aspherical mass-loss from the
central star is one of the favoured scenarios \citep[e.g.,][]{Dwarkadas2002,Garcia-Segura2005}.
In a radiatively driven wind, rotation 
may result in latitudinal variations in the wind, with high velocities and 
larger wind-momentum gas at the poles compared to the equator.
Rotation can also significantly affect the magnetic channelling
of a hot star wind  \citep[e.g.,][]{Ud-Doula2008}. 

A pivotal issue in the role of stellar rotation in PNe $-$ and the subject 
of this Letter $-$ is whether the precursor (AGB) and remnant (CSPN) 
rotation is fast enough to produce the contrasting equatorial and polar 
outflows.
Observational constraints on AGB rotation rates suggest upper limits
of a few km s$^{-1}$  \citep[e.g.,][]{deMedeiros1999},
though \citet{Ignace1996} argue that very modest rotation rates in AGB stars
are not likely to result in aspherical geometries.
In their study of the evolution of rotating stars from main-sequence
to white dwarf, \citet{Suijs2008} predict that single star progenitors
of CSPNe have very little or no rotation at all in this phase.

In this Letter we provide new perspectives and constraints on the rotation 
of CSPN by studying the temporal evolution of large-scale (coherent) wind 
structure, which we argue is rooted at the stellar surface. Over the past 
two decades extensive UV spectral time-series observations of OB and WR 
stars, obtained primarily with the {\it International Ultraviolet Explorer} 
($IUE$) satellite, have revealed systematic behaviour of wind variability. 
The most thoroughly studied objects exhibit cyclical variation in the stellar 
winds that are causally connected to the photosphere, with the time-scales of 
the variability related to the rotation periods of the massive stars 
\citep[e.g.,][]{Massa1995,Fullerton1996,Kaper1996,deJong2001,Prinja2002,Chene2011}.
The physical interpretation of this rotational modulation is the action of spiral-shaped 
structure in the wind flow caused by perturbations at the base of the wind 
and carried by stellar rotation.  Using {\it Far-Ultraviolet Spectroscopic 
Explorer} ($FUSE$) data, similar wind structure has been demonstrated to 
exist in the fast winds of CSPN \citep[e.g.,][]{Prinja2007,Prinja2012}.

In this Letter we report on high-resolution, high-quality {\it Hubble Space 
Telescope} ($HST$) time-series spectroscopy of the central star of the 
hydrogen-rich (Cat's Eye) nebula NGC~6543.  The time 
dependence of the fast wind of NGC~6543 was previously shown to exhibit 
discrete absorption components, which traverse the absorption troughs of 
unsaturated wind lines on time-scales $\lesssim 1$ hour \citep{Prinja2007}.  
We present here a multi-line study of the variability in the \hst\ 
spectra. The clarity of the structure seen in the resonance, excited state 
and iron UV lines provides key details on the nature of the wind structure, 
and thus its origins and connection to the rotation of the central star.

\section{{\it HST}/STIS observations}
The central star (HD~164963) of the well studied PN NGC~6543 was observed 
with \hst\ for Program \#12489 (D.\ Massa PI) on 2012 May 16.  Repeated 
Space Telescope Imaging Spectrograph (STIS) E140M grating (spectral 
resolution $\simeq 10$ $\kms$) observations were secured for 5 contiguous 
\hst\ orbits in ACCUM mode and with individual exposure times of 180-s.  
A total time-series of 95 spectra were secured spanning $\sim$ 5.8 hours, 
each with a wavelength range $\sim$ $\lambda\lambda$1140 to 1710 {\AA}. 
The signal-to-noise resolution per element of an individual spectrum 
varies from $\sim$ 5 at the extreme wavelength ends to $\sim$ 30 at
1300{\AA}.

The rich UV spectrum of HD~164963 contains several key diagnostic lines 
including the resonance line doublets of N{\sc v} $\lambda\lambda$1239, 
1243, Si{\sc iv} $\lambda\lambda$1394, 1403, and C{\sc iv} $\lambda 
\lambda$1549,1551; the excited state wind lines of O{\sc iv} $\lambda 
\lambda$1339, 1344 and O{\sc v} $\lambda$1371; 
He{\sc ii} $\lambda$1640, as well as 'forests' of Fe{\sc v} and Fe{\sc vi} 
absorption lines.  Table~1 lists the fundamental parameters of HD~164963 
we adopt for this paper.

\begin{table}
\begin{center}
\caption{Adopted parameters for the central star of NGC~6543}
\begin{tabular}{ll}
\hline\hline
Parameter & Value \\
\tableline
Spectral type & Of(H-rich) \\
Luminosity & 1585$L_\odot$ \\
$T_{\rm eff}$ & 67000 K \\ 
Radius & 0.3R$_\odot$ \\
Radial velocity & $-$66 km s$^{-1}$ \\
Terminal velocity & 1340 km s$^{-1}$ \\
Velocity law ($\beta$) &  2   \\
\hline
\end{tabular}

\tablecomments{Parameters are from the non-LTE analysis of Georgiev et al. 2008}
\end{center}
\end{table}

\begin{figure}
\epsscale{1.0}
\plotone{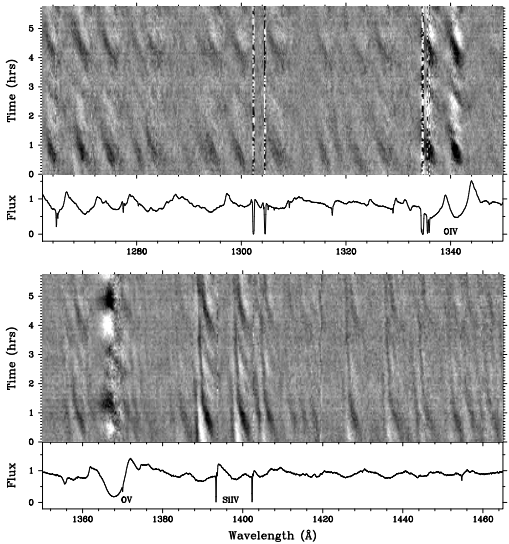}
\caption{Dynamic spectra revealing evidence for systematic variability in 
almost every detectable absorption line in the \hst/STIS spectrum of the 
central star of NGC~6543.  Recurrent large-scale wind structure is 
diagnosed by the Si{\sc iv} resonance doublet and O{\sc iv} and O{\sc v} 
excited state lines.  Sympathetic variations are also evident in numerous
Fe{\sc vi} ($\lambda\lambda$1260 to 1310) and Fe{\sc v} 
($\lambda\lambda$1390 to 1460) lines.\label{fig1}}
\end{figure}

\section{Evidence for evolving large-scale wind structure}
Previous studies established that the fast wind of NGC~6543 is variable on 
short ($\sim$ hourly) time-scales, with diagnostic changes evident in 
optical spectroscopy \citep{Mendez1990}, photometry \citep{Bell1994} 
and FUV to UV spectroscopy \citep{Prinja2007,Patriarchi1997}.  
The widespread and rapid line profile variation present in our 
\hst/STIS time-series is demonstrated in Fig.\ 1 which shows ratios of 
the individual STIS spectra divided by the mean for the time-series as a 
dynamic spectrum (i.e. a two-dimensional image of the time ordered 
ratios). 

The temporal behaviour seen in Fig.\ 1 is remarkable.  In addition to the 
`classic' blueward migrating and recurring discrete absorption components 
(DACs) present in the unsaturated resonance line doublet of Si{\sc iv} 
$\lambda\lambda$1394, 1403 and in the subordinate excited state lines of 
O{\sc iv} $\lambda\lambda$1339, 1344 and O{\sc v}$\lambda$1371, the 
numerous Fe{\sc vi} absorption features between $\sim$ $\lambda\lambda$1260 
to 1310{\AA} and Fe{\sc v} lines between $\lambda\lambda$1390 to 1460{\AA} 
also vary {\it in concert} with the wind lines.  Three DAC episodes are 
evident over $\sim$ 5.8~hr (see also Fig.\ 2): There is some empirical 
resemblance in optical depth and velocity dispersion between the first 
episode (T $\sim$ 0.3 to 2~hr) and the third feature (T $\sim$ 3.7 to 
5.5~hr).  The second DAC episode (T $\sim$ 2.3 to 3.3~hr) is weaker though 
it has a similar blueward velocity progression.
We note that the individual spectra were normalized to suppress an overall
continuum variability of $\sim \pm$4\%, which is similar in amplitude and
time scale to that seen in the optical by \citet{Bell1994}.
Because the
time dependence is not obviously in phase with the DACs, its exact origin
is not clear.
However, the wavelength independence of the continuum variability
suggests that it results from some sort of geometric obscuration instead
of temperature irregularities.

The reaction of the deep-seated Fe lines indicates that the wind structures 
originate very close to or at the stellar surface.  The variations in 
excited lines of O{\sc iv} and O{\sc v} provide further strong support of 
this deep-seated origin.  In general, a stellar flow need not be a 
monotonically increasing function of distance from the star and an 
absorption feature seen at low velocity in a resonance line need not 
originate close to the star \citep[e.g.,][]{Owocki1988}.  In 
contrast, in dense expanding winds the populations of the excited levels 
require a strong EUV radiation field, thus these transitions have a strong 
dependence on the radial distance from the star.  This means that large 
populations of the lower levels of O{\sc iv} and O{\sc v} ions can only 
occur close to the star.  We conclude therefore that the evolving features 
seen in Fig.~1 in the fast wind NGC~6543 originate at the stellar surface.

The short time-scale fluctuations in the numerous Fe 'forest' lines also 
imply that the surface conditions on the CSPN are non-uniform, and likely 
to affect the results of 1-D model atmosphere analysis.  This may account 
for the inconsistencies \citet{Georgiev2008} report in their non-LTE 
analysis of HD~164963, including an overabundance of Si, observed spectral 
features not present in the model, differing wind velocities for different 
resonance lines, and the need for additional ionization sources in the 
external part of the fast wind.

\section{Co-rotating interaction regions}

The variability patterns observed in O{\sc iv}, O{\sc v} and He{\sc 
ii}~$\lambda$1641 are shown in closer (velocity) detail in Fig.\ 2, where 
some further characteristics are revealed.  Firstly, the line profile 
modulations tend to be bow-shaped in the direction of increasing time, 
i.e., an absorption enhancement occurs first at an intermediate velocity 
and then spreads to higher (bluer) and lower (redward) velocities 
simultaneously at a later time.  This `phase bowing' is particularly 
evident in the dynamic spectra of O{\sc iv} and He{\sc ii}.  Similar forms 
of variability have been identified in the stellar winds of OB stars \citep[e.g.,][]{Fullerton1997,deJong2001,Prinja2002}.  
Hydrodynamic models by \citet{Cranmer1996} \citep[see also][]{Lobel2008} 
showed that large scale, pro-azimuthal, wind structure in the 
form of co-rotating interaction regions (CIRs) can explain the 
characteristics of the bowed discrete absorption components.  The CIR 
structures are akin to those seen in the solar wind \citep[e.g.,][]{Hundhausen1972,Mullan1984}, 
where different sectors of the stellar surface give rise to 
winds which accelerate differently.  When, due to rotation, these different 
sectorial flows interact, an interface with a spiral pattern results and a 
velocity plateau develops along the interface.  As the CIR pattern rotates 
across our line of sight to the stellar disk, it creates an absorption 
in the dynamic spectrum which evolves toward high velocity.  Intensity 
irregularities at the stellar surface, such as spots, non-radial 
pulsations, or granulation can potentially alter the radiative 
wind acceleration locally, thus giving rise to streams of faster and slower
material, which interact to form the CIRs.

We also note in Fig.\ 2 evidence for an ionization shift, where the 
absorption enhancements with respect to the mean profile (i.e. darker 
patches) in O{\sc v} lag those in the O{\sc iv} and He{\sc ii} by about 
1~hr.  (See, e.g., the episode between $\Delta$T $\sim$ 4 and 5-hr in Fig.\ 
2).  This may be evidence for an ionization gradient across the CIR, with 
the lower ions on the leading edge of the CIR and higher ions on the 
trailing edge.

\begin{figure}
\epsscale{1.0}
\plotone{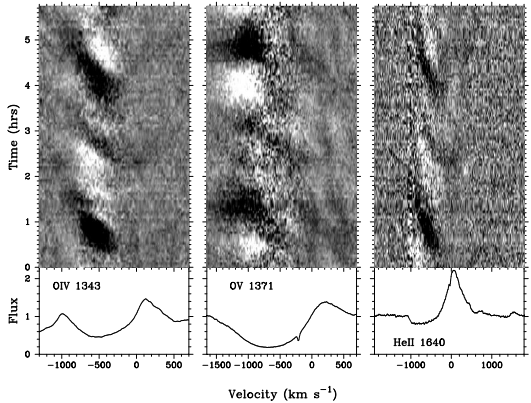}
\caption{Dynamic spectra of 
O{\sc iv}, O{\sc v} and He{\sc ii} revealing bow-shaped
modulation that extends blueward and redward simultaneously
in time. Note also the ionization shift between
O{\sc v} and the lower ions O{\sc iv} and He{\sc ii}.\label{fig1}}
\end{figure}

\section{Constraining the stellar rotation rate}\label{rot_rate}
We have demonstrated that the UV wind lines of NGC~6543 are extensively 
variable, and that the variability results from large-scale, coherent wind 
structure.  The characteristics of the variability, including bow-shaped 
modulations, are consistent with co-rotating interaction regions (CIRs) 
that are rooted at the stellar surface.  In this section we use the 
temporal evolution of the DACs to constrain the rotation rate of the 
central star in NGC~6543.

By their very nature, CIRs infer rotation.  They have been detected in the 
winds of O  \citep{Kaper1997,deJong2001} B \citep{Fullerton1997,Prinja2002} and WR stars \citep[e.g.,][]{Chene2011}.  
Typically, the signatures of 2 strong CIRs are seen, but these are 
sometimes accompanied by two weaker arms (Fullerton et al. 1997, de Jong et 
al. 2001).  It seems reasonable, therefore, to associate the two strong features 
which occur at $\Delta t \simeq 0.3$ and 3.7~hours in Figures 1 \& 2 as 
strong CIRs and the one near 2.3~hours as a weaker one.  Thus, we estimate 
the rotation period of HD~164963 is $2 \times (3.7 - 0.3) \sim 6.8$ hours.  In 
this case, its rotation rate is $\simeq 179 R$~km~s$^{-1}$, where $R$ is
the stellar radius in 
solar units.
The most recent line-blanketed non-LTE line synthesis analysis of
the central star yields
estimates of $R$ for HD~164963 that vary between 0.3 (Table~1) and 
0.6 (Herald and Bianchi 2011).  We therefore estimate its rotation rate to be in 
the range $54 \leq v_{rot} \leq 108$~$\kms$.  

\section{Discussion}

State-of-the-art stellar evolution calculations of rotating, single low-mass stars predict 
a substantial spin-down during their late phases of evolution. This is because, despite a 
possibly high initial rotation rate, the dramatic expansion and increased mass-loss during 
the RGB and AGB phases result in huge angular momentum loss. Coupling between the core and 
the envelope is provided by angular momentum transport processes, arising from rotational 
instabilities and magnetic torques. This coupling is required to explain the slow rotation 
rate of white dwarfs and, for more massive progenitors, neutron stars \citep{Heger2005,Suijs2008}. 
Moreover an angular momentum transport in radiative zones like the one 
provided by magnetic torques is necessary to reproduce the near uniformity of the solar 
rotation profile \citep[e.g.,][]{Eggenberger2005} and possibly the recent asteroseismic observations of 
the red giant KIC 8366239 \citep{Beck2012,Eggenberger2012}.
However we want to stress that the angular momentum transport
produced by magnetic torques relies on the existence of a dynamo in the
radiative layers of stars \citep[Tayler-Spruit dynamo,][]{Spruit2002},
which is currently debated  \citep[e.g.,][]{Zahn2007}. Other angular momentum transport
processes in radiative zones have also been proposed \citep[e.g., gravity waves][]{Talon2005}.

We have used the Modules for Stellar Experiments in Astrophysics \citep[Mesa version 4298][]{Paxton2011} code to calculate the evolution of a grid of low-mass stars (1.5$\mso$, 
2$\mso$ and 3$\mso$) with metallicity Z = 0.02 and initial equatorial rotational velocity 
of 250$\kms$. This code accounts for the transport of angular momentum by rotational 
instabilities and magnetic torques \citep[see e.g.,][]{Heger2000,Heger2005,Suijs2008}. 
We adopted the mass-loss rate of \citet{Reimers1975} during the RGB evolution and the 
one of \citet{Bloecker1995} during the AGB and post-AGB phase. Our 1.5$\mso$ model provides the 
best fit to the values of luminosity (1585$\lso$) and effective temperature
(67000K) determined by \citet{Georgiev2008}, however the major results discussed below do 
not change for the other calculations. We find that when the star reaches the CSPN phase, 
it has lost most of its angular momentum and its surface rotation velocity is much lower 
than $1\kms$ (see Fig. 3).
Therefore calculations that adopt the physical ingredients required to explain both 
progenitor and descendant stars of CSPN, are at odds with the limits on the rotation rate 
of NGC~6543 derived in Sect. 5. Note that DACs seem to be frequent among CSPN 
\citep{Prinja2012}. Since to be observable DACs usually need to recur on a short period, 
this suggests that rapid rotation could be widespread among CSPN (where here rapid means 
rotating faster than the predictions of single stellar evolution calculations).
 
An interesting possibility is that NGC~6543 did not evolve as a single star. In a binary 
system spin-up of one or both components could result from mass accretion, tidal 
interaction or a merger. This is particularly interesting in the case of CSPN, as binarity 
has been often discussed as a possible explanation for the asymmetric shapes of PNe \citep[see e.g.,][]{Iben1995,Balick2002,Huarte-Espinosa2012}.

\begin{figure}
\epsscale{1.25}
\plotone{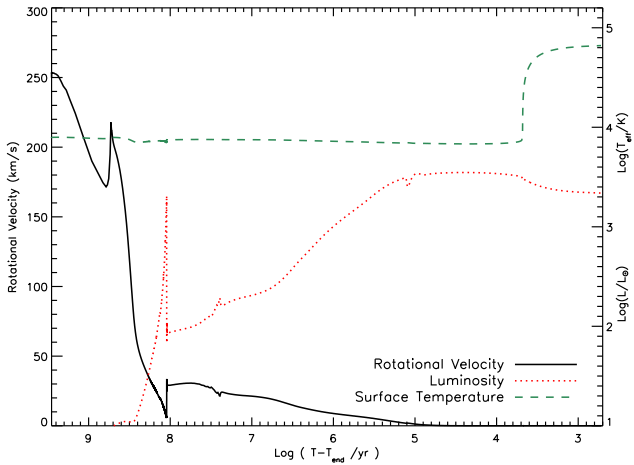}
\caption{Evolution of the equatorial rotational velocity for our 1.5$\mso$
model with metallicity Z = 0.02 and initial equatorial rotational velocity
of 250$\kms$. The horizontal log scale shows the time to the end of the calculation, when the
luminosity and surface temperature roughly correspond to the estimated values for the central star of
NGC~6543 (See Table 1).
The model includes angular momentum transport induced by rotational instabilities and magnetic torques.
The evolution of stellar luminosity (dotted line) and surface temperature (dashed line) are also shown for
reference.
\label{fig3}}
\end{figure}

CSPN are hot and luminous, and due to an iron opacity peak at about 150000 K, they can 
develop a subsurface convection region. This occupies a negligible fraction of the stellar 
mass, but at the same time occurs very close to the stellar surface. Our model of the 
central star of NGC~6543 reveals the presence of such convective layer, with convective 
velocities of about $10\kms$ ($\sim$15\% of the local sound speed; CSPN models with higher 
initial mass result in higher velocities). Note that the presence of this convective layer 
is a robust result. It is a structural property depending only on the opacity and 
luminosity, and therefore we do not expect major changes if the star went through binary 
interactions. Like the case of massive OB stars, it is possible that such convection zones 
can produce density and velocity fluctuations at the surface, stochastically excite 
oscillations and produce magnetic fields \citep{Cantiello2009, Cantiello2011}. 
Interestingly the recent observation of stochastic brightness variations in the CSPN 
of NGC~6826 \citep{Jevtic2012} could be related to the presence of such subsurface 
convective layers. The non-uniform surface conditions revealed by the
Fe{\sc V} and Fe {\sc VI} line 
profile variability in NGC~6543 seem to further support this notion. If a 
dynamo\footnote{The convective turnover timescale is of order hours, comparable to the 
inferred rotational period of NGC~6543. This means that the Rossby number is of order unity 
and that an $\alpha\omega$-dynamo might be at work.} is at work in the iron convection zone 
of NGC~6543, 
adopting the densities and MLT convection velocities from our models
one finds an equipartition field B$\sim$2kG (inside the convective region). 
Similar to the case of massive OB stars discussed in
\citet{Cantiello2011}, these 
fields can in principle reach the stellar surface through magnetic buoyancy and affect the 
stellar surface (e.g. creating bright spots). Bright spots have been discussed as a way to 
produce CIRs \citep{Cranmer1996}, which might provide an 
explanation for the common emergence of DACs in CSPN \citep{Prinja2007, Prinja2012}.

Our work shows that observations of DACs in CSPN can provide important constraints on the 
surface rotational velocity of these intriguing stars. In turn this information can be used 
to constrain their evolutionary history, potentially shedding light on the puzzling origin 
and shaping of planetary nebulae.

\vskip 0.5cm
\acknowledgments
MC would like to thank Bill Paxton for useful discussions.
Support for this work was provided by NASA through grant number 
HST-GO012489 from the Space Telescope Science Institute, which is operated 
by AURA, Inc., under NASA contract NAS5-26555.
This research was supported in part by the National Science Foundation under Grant No. NSF PHY05-51164.

\bibliographystyle{apj} 
\bibliography{ref}

\end{document}